# Title: The magnetic field and turbulence of the cosmic web measured using a brilliant fast radio burst

**Authors:** V. Ravi[1]*[†], R. M. Shannon[2,3]*[†], M. Bailes[4,5], K. Bannister[2], S. Bhandari[4,5], N. D. R. Bhat[3,5], S. Burke-Spolaor[6], M. Caleb[7,4,5], C. Flynn[4,5], A. Jameson[4,5], S. Johnston[2], E. F. Keane[8], M. Kerr[2], C. Tiburzi[9], A. V. Tuntsov[10], H. K. Vedantham[1]

**Affiliations:**

[1]Cahill Center for Astronomy and Astrophysics, MC249-17, California Institute of Technology, Pasadena, CA 91125, USA.

[2]CSIRO Astronomy and Space Science, Australia Telescope National Facility, P.O. Box 76, Epping, NSW 1710, Australia.

[3]International Centre for Radio Astronomy Research, Curtin University, Bentley, WA 6102, Australia.

[4]Centre for Astrophysics and Supercomputing, Swinburne University of Technology, P.O. Box 218, Hawthorn, VIC 3122, Australia.

[5]ARC Centre of Excellence for All-sky Astrophysics (CAASTRO).

[6]National Radio Astronomy Observatory, Array Operations Center, P.O. Box 0, Socorro, NM 87801-0387, USA.

[7]Research School of Astronomy and Astrophysics, Australian National University, ACT 2611, Australia.

[8]SKA Organisation, Jodrell Bank Observatory, SK11 9DL, UK.

[9]Max-Planck-Institut für Radioastronomie, Auf dem Hügel 69, 53121 Bonn, Germany

[10]Manly Astrophysics, 3/22 Cliff Street, Manly, NSW 2095, Australia.

*Correspondance to: vikram@caltech.edu, Ryan.Shannon@csiro.au

[†]These authors contributed equally to the work.

**Abstract**: Fast radio bursts (FRBs) are millisecond-duration events thought to originate beyond the Milky Way galaxy. Uncertainty surrounding the burst sources, and their propagation through intervening plasma, has limited their use as cosmological probes. We report on a mildly dispersed (dispersion measure 266.5±0.1 pc cm$^{-3}$), exceptionally intense (120±30 Jy), linearly polarized, scintillating burst (FRB 150807) that we directly localize to 9 arcmin$^2$. Based on a low Faraday rotation (12.0±0.7 rad m$^{-2}$), we infer negligible magnetization in the circum-burst plasma and constrain the net magnetization of the cosmic web along this sightline to <21 nG, parallel to the line-of-sight. The burst scintillation suggests weak turbulence in the ionized intergalactic medium.

**One Sentence Summary:** Cosmological radio bursts of remarkable brightness uniquely probe aspects of the ionized content of the Universe.

**Main Text:**

A recently recognized population of fast radio bursts (FRBs) of likely extragalactic origin (*1, 2*) may revolutionize astrophysics and cosmology. Besides probing a heretofore-unknown astrophysical phenomenon, the bursts potentially carry imprints of propagation through inhomogeneous, magnetized plasma in the ionized interstellar media of other galaxies, and the diffuse intergalactic medium (IGM). Simultaneous measurements of redshifts and line-of-sight free electron column densities for FRBs can constrain the cosmological mass-density and ionization history of baryons (*3-5*).

We detected FRB 150807 with the 64-m Parkes radio telescope, using the 21-cm multibeam receiver (*6*), while conducting timing observations of the millisecond pulsar PSR J2241−5236. This receiver is sensitive to 13 overlapping regions (beams) on the sky within a 2° diameter circle. We performed a real-time search for FRBs commensally with the pulsar timing experiment, using standard hardware (*7*) and search techniques (*8-10*).

The properties of FRB 150807 are given in Table 1, and calibrated burst data are shown in Fig. 1. The burst was detected in two adjacent beams. We use this two-beam detection and a model of the multibeam response (*10*) to constrain the position of the burst to a 9 arcmin$^2$ region with 95% confidence (Fig. 2), and correct the flux density and spectral shape of the burst due to the telescope response relative to its off-axis localization. Over the 1182 – 1519.5 MHz band, the mean burst fluence is 50±20 Jy ms and the bandwidth-averaged peak flux density is 120±30 Jy, with the uncertainties dominated by the telescope response at the burst position. Direct fluence measurements have not been possible for previous FRBs.

The line-of-sight free electron column density for FRB 150807, measured in units of dispersion measure (DM), is 266.5±0.1 cm$^{-3}$ pc. This substantially exceeds the expected foreground Milky Way DM, predicted to be 70±20 cm$^{-3}$ pc along the burst sightline (Galactic longitude $l$ = 336.71±0.03°, latitude $b$ = −54.40±0.03°), which includes contributions from ionized plasma in both the Milky Way disk [40±20 cm$^{-3}$ pc (*12-14*)] and halo [30±10 cm$^{-3}$ pc (*15*)] (*16*). The DM of FRB 150807 is the smallest hitherto reported for an FRB.

The spectrum of FRB 150807 is strongly enhanced between 1250 and 1300 MHz (Fig. 1 D). This is reminiscent of the spectra of bursts from the repeating FRB 121102 (*17*), and may be intrinsic to the source. The enhancement could also be due to diffractive scintillation caused by scattering in the Milky Way: the expected scintillation bandwidth induced by Milky Way plasma density fluctuations is expected to be >8 MHz at these frequencies (*12*). The transient nature of scattering-induced magnifications (*18*) may explain the general absence of repeated bursts, if FRBs represent frequent bursts from compact objects (*19-21*). However, it is also possible that our follow-up observations, totaling 215 hr (*10*), were insufficiently sensitive to detect repeat bursts.

The localization of FRB 150807 can be used to estimate the distance at which it was emitted, if we can associate the FRB with a star or a galaxy. The deepest archival images of the sky-localization area (Fig. 2) contain nine objects brighter than a *Ks*-band magnitude of 19.2 (*11*):

three stars and six galaxies. The brightest galaxy is at a distance between 1 and 2 Gpc (95% confidence) estimated from its photometric redshift (supplementary text section S2, *10*). The other galaxies are factors of >6 fainter than the brightest. Through a comparison of their infrared magnitudes with empirical and theoretical distributions of galaxy luminosities at different distances, they are all expected to be >500 Mpc distant (*30*). Additionally, a similar analysis yields a low probability (<5%) of a fainter undetected galaxy with a mass >$10^9$ solar masses closer than 500 Mpc. If FRB 150807 originated in a galaxy, a distance of >500 Mpc is therefore likely unless the galaxy was <$10^9$ solar masses.

The 80±1% linear-polarization fraction of FRB 150807 enabled us to measure the Faraday rotation induced by magnetization of the dispersing plasma. The pulse-averaged rotation measure (RM) is 12.0±0.7 rad m$^{-2}$ (*10*). The nearby millisecond pulsar PSR J2241−5236, located 0.5° away on the sky, has an RM of 13.3±0.1 rad m$^{-2}$ despite a low DM of 11.41±0.01 cm$^{-3}$ pc (*22*). The Milky Way contributions to extragalactic source RMs vary by <~10 rad m$^{-2}$ on degree angular scales at high Galactic latitudes (*23-25*). The pulsar RM is consistent with expectations for the Milky Way contribution along this sightline (*24, 25*); we hence adopt it as such. We constrain the net extragalactic line-of-sight magnetic field, $<B_\parallel>$, weighted by the free electron density distribution and likely incorporating numerous reversals of polarity, to be $<B_\parallel>$ < 21(1+$z_{mean}$) nG (>95% confidence), where $z_{mean}$ is the mean redshift of the intervening electron-density distribution (*26*), expected to be ~0.1 if the burst is at a Gigaparsec distance. The combined DM and RM measurements directly constrain the Gigaparsec-scale magnetization of the cosmic web along this sightline. Our results are consistent with numerous models (*26, 27*) that predict that at most nano-Gauss magnetic fields pervade the cosmic web. Additionally, FRB progenitor theories that propose emission from young neutron stars or other objects embedded in highly magnetized star forming regions or galaxy centers (*19-21*) may be inconsistent with the low RM of FRB 150807, unless the net magnetization is externally cancelled.

The dispersion-corrected burst dynamic spectrum (Fig. 1 C) shows strong modulations in both frequency and time below the instrument resolution, which are inconsistent with thermal noise associated with the telescope or the burst (supplementary text section S1, *10*). Instead, the intensity variations, portions of which exceed 1 kJy, have an exponential distribution. As the burst width at each frequency is consistent with the amount of dispersion in each spectrometer channel, implying that the burst is temporally unresolved, the temporal structure in the dynamic spectrum corresponds to spectral features that are narrower than the channel width, which are at random frequencies within the channels and therefore incorrectly de-dispersed. We estimate the characteristic bandwidth of these spectral features to be 100±50 kHz.

The spectral features with exponentially distributed intensities imply that the burst temporal profile is dominated by structure on few-microsecond scales; if the burst width was 10 μs, the measured fluence would imply a mean flux density of 5 kJy. This structure could be intrinsic to the burst emission mechanism, in which case FRB 150807 would be similar to giant pulses from the Crab pulsar (*28*). In this interpretation, we would expect to observe similar spectral intensity variations in other FRBs that were detected with high significance, unless other FRBs have substantially different pulse durations. The 100 kHz structure in the spectrum of FRB 150807 may instead represent diffractive scintillation, caused by the scattering of the burst in inhomogeneous plasma along the line of sight with a characteristic timescale $\tau_d = 1 / (2\pi \times 100$ kHz) = 1.6 μs (*18*). By analyzing the multi-frequency burst profile, we place an independent upper limit on $\tau_d$ of 27 μs (95% confidence) at 1.3 GHz (supplementary text section S1.4, *10*). FRB 150807 thus exhibits weaker scattering than measured for any other FRB.

Scattering strength, quantified by the characteristic delay ($\tau_d$) of scattered rays, is related to the line-of-sight integral of the plasma-density fluctuation power spectrum [the scattering measure (SM)], under specific assumptions about the location and geometry of the scattering medium (supplementary text section S1.3, *10*). Contributions to the burst DM, RM and SM can be made by plasma within the host galaxy (if the burst originated within a galaxy), within the Milky Way, and in the IGM. The 100 kHz structure is much narrower in frequency than the expectation for Milky Way scintillations (>8 MHz) along this sightline (*12*). For the maximum possible redshift (0.4, >95% confidence) of FRB 150807 implied by its DM (*15*), intervening galaxies have only a few percent probability of being present (*29*), and would imply $\tau_d \gg$ 1 ms (*30*). If $\tau_d$ =1.6 μs, the low RM of the burst suggests that the scattering is unlikely to have originated in a disk-galaxy host system regardless of the host DM contribution. In Fig. 3, we compare estimates for FRB scattering strengths to those for Milky Way pulsars. We apply the standard thin-screen model for the scattering geometry, which relates the SM to $\tau_d$ as

$$\mathrm{SM} = 2.73 \times 10^{17} \, (1 + z_{scr})^{17/6} \, (\nu \,/\, 1.3 \text{ GHz})^{11/3} \, [\tau_d \, (D_{eff} \,/\, 1 \text{ Gpc})^{-1}]^{5/6} \text{ m}^{-17/3} \qquad (1)$$

where $\nu$ is the observing frequency, $z_{scr}$ is the redshift of the screen, and $D_{eff} = (D_{scr} / D)(D - D_{scr})$, where $D_{scr}$ is the distance from the Earth to the scattering screen, and $D$ is the Earth-source distance (*33*). If FRB 150807 originated in a disk galaxy like the Milky Way, the magnitude of its RM (|RM|) would be unexpectedly low for its SM if $\tau_d \geq$ 1.6 μs. Also, depending on whether or not the host dominated the DM, the |RM| would either be unexpectedly low for the DM, or the SM would be unexpectedly high for the DM, respectively.

As suggested for other FRBs (*30*), it is possible that the 100 kHz structure in FRB 150807 represents weak scattering in a host galaxy that dominated the dispersion, with lower-than-expected |RM| and SM relative to Milky Way disk sightlines. The host would be near to the Milky Way, and the localization implies that the galaxy would likely have a low mass (<~$10^9$ solar masses). It is also possible that the dispersion of FRB 150807 was dominated by the IGM contribution, which would suggest a distance $D \sim$ 1 Gpc (*15*), in good agreement with the localization-based constraint. If we model the IGM as a homogeneous scattering medium along the entire sightline, the SM is approximately given by setting $D_{eff} = D/2$. Setting $D = $ 1 Gpc, and assuming $\tau_d$ = 1.6 μs, we find SM = $9.5 \times 10^{12}$ m$^{-17/3}$. Although this is consistent with a prediction of SM ≈ $4 \times 10^{12}$ m$^{-17/3}$ in the IGM for sources at this distance (*33*), such a high SM may be inconsistent with the observed thermal stability of the IGM (*34*).

Despite its unique properties, FRB 150807 is not excluded from being drawn from the same source population as the majority of FRBs. In this case, a comparison between FRB 150807 and other FRBs provides support for the cosmological origin of a large fraction of the FRBs. The event with the second-lowest DM, FRB 010724 (*1*), was detected in four beams of the Parkes multibeam system, implying a comparable fluence to FRB 150807. Therefore, DM may correlate with source distance among FRBs, which would only be expected for a cosmological population with sub-dominant host-galaxy contributions. The discovery of FRB 150807 also partially resolves the lack of low-DM FRBs, which has been suggested as evidence for a nearby population with host-dominated DMs (*30, 35*). The weak scattering of FRB 150807 relative to other FRBs with detected scattering could indicate a relation between dispersion and scattering strength characteristic of a clumpy IGM, similar to that observed in the Milky Way interstellar medium (*36*). Finally, an analysis of the FRB fluence distribution, including FRB 150807, finds evidence against a local-Universe population (*37*). This analysis also predicts that events with fluences similar to FRB 150807 (that is, >50 Jy ms) are not rare, with a rate at ~1.3 GHz of 190±60 sky$^{-1}$ day$^{-1}$.

**Acknowledgments:** We thank both D. Schnitzeler and the SUPERB collaboration for useful feedback on the manuscript. The Parkes radio telescope is part of the Australia Telescope National Facility which is funded by the Commonwealth of Australia for operation as a National Facility managed by Commonwealth Science and Industrial Research Organization (CSIRO). The data for FRB 150807 reported in this paper are made available through the CSIRO data access portal (http://doi.org/10.4225/08/580ffcb0d68cc).

**Supplementary Materials**
www.sciencemag.org
Materials and Methods
Supplementary Text
Figs. S1 – S10
Tables S1 – S4
References (38 – 78)
Supplementary computer code (gen_maps.py, comalobes.py).

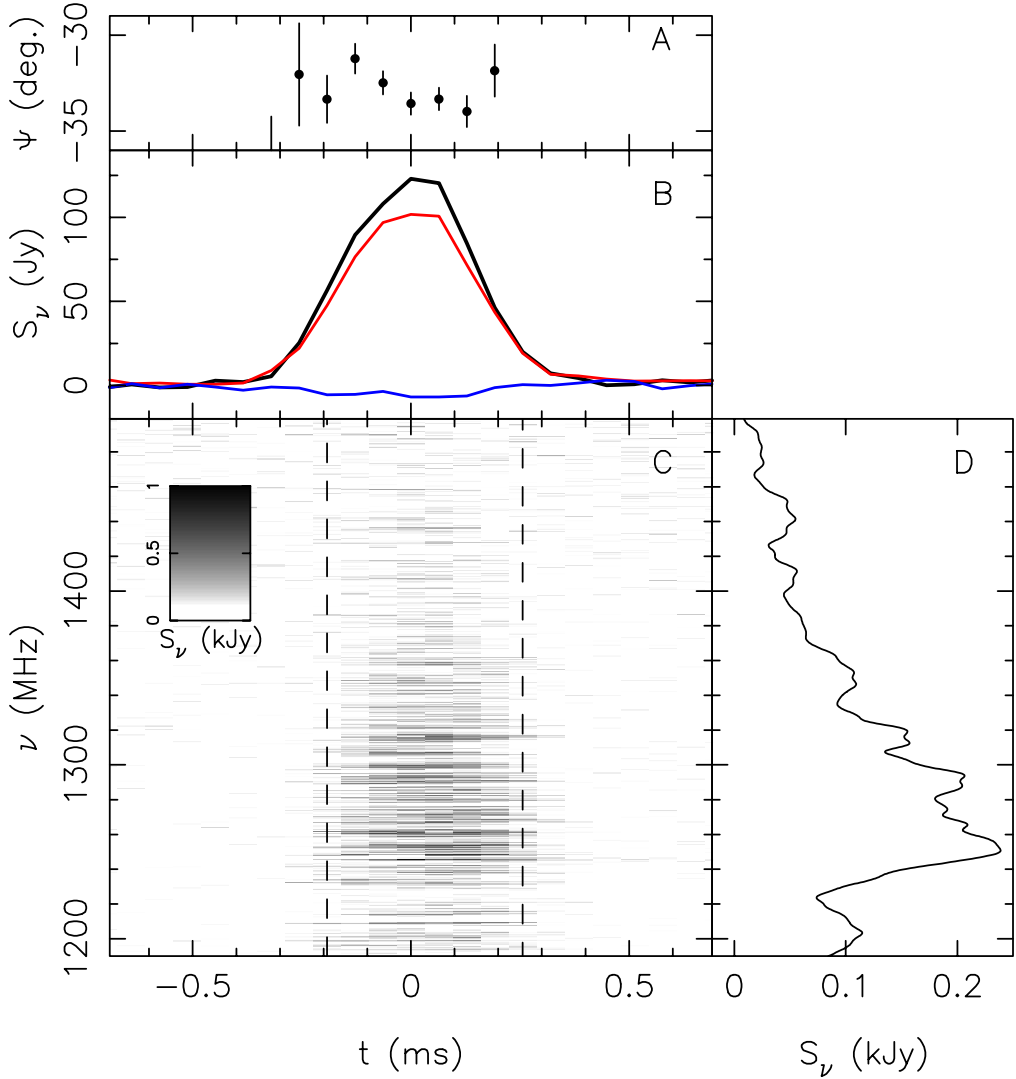

**Fig. 1**. **Polarization and spectral properties of FRB 150807.** *Panel A*: Absolute position angle (ψ) of the electric field polarization vector. *Panel B*: Total intensity (black), linear polarization fraction (red), and Stokes V (blue) time-series profiles of the burst, averaged over all frequency channels. *Panel C*: De-dispersed dynamic spectrum of the burst. The intensity scale is indicated by the inset color bar. The small bars on the left of the plot show frequency channels removed because they contained radio-frequency interference. *Panel D*: The time-averaged spectrum of the burst, smoothed with a 5 MHz Gaussian filter. The time-resolution of the data in Panels A, B, and C is 64 µs, and the frequency-resolution of the data in Panels C and D is 390.625 kHz. The data displayed in all panels have been corrected for the frequency dependent instrumental gain, the instrumental response to a polarized source, and the effects of the off-axis burst position. Additionally, the data displayed in panels B, C, and D have had the mean off-pulse levels subtracted from each channel.

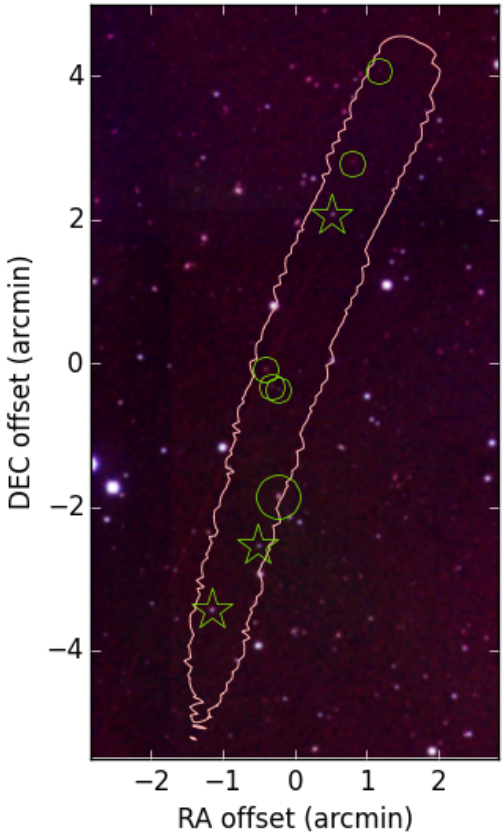

**Fig. 2. Containment region and possible hosts of FRB 150807.** The pink trace shows the 95% confidence containment region from our modeling of the Parkes multibeam response pattern. Besides an 11 arcsecond uncertainty in the Parkes telescope pointing, the size of the error region is dominated by the uncertainty in the multi-frequency signal to noise ratio measurements in the weaker of the detected beams. The background is a three-color composite of images in the *J*, *H*, and *Ks* near-infrared bands from the Visible and Infrared Survey Telescope for Astronomy (VISTA) Hemisphere Survey (VHS, Ref. *11*). The image is centered on right ascension 22:40:23.04 and declination -53:16:12.4 (J2000). The stars and circles represent sources identified as Milky-Way stars and galaxies respectively (supplementary text section S2, *10*). A large circle highlights the position of VHS7, the brightest galaxy inside the containment region, which has a distance of between 1 and 2 Gpc (95% confidence).

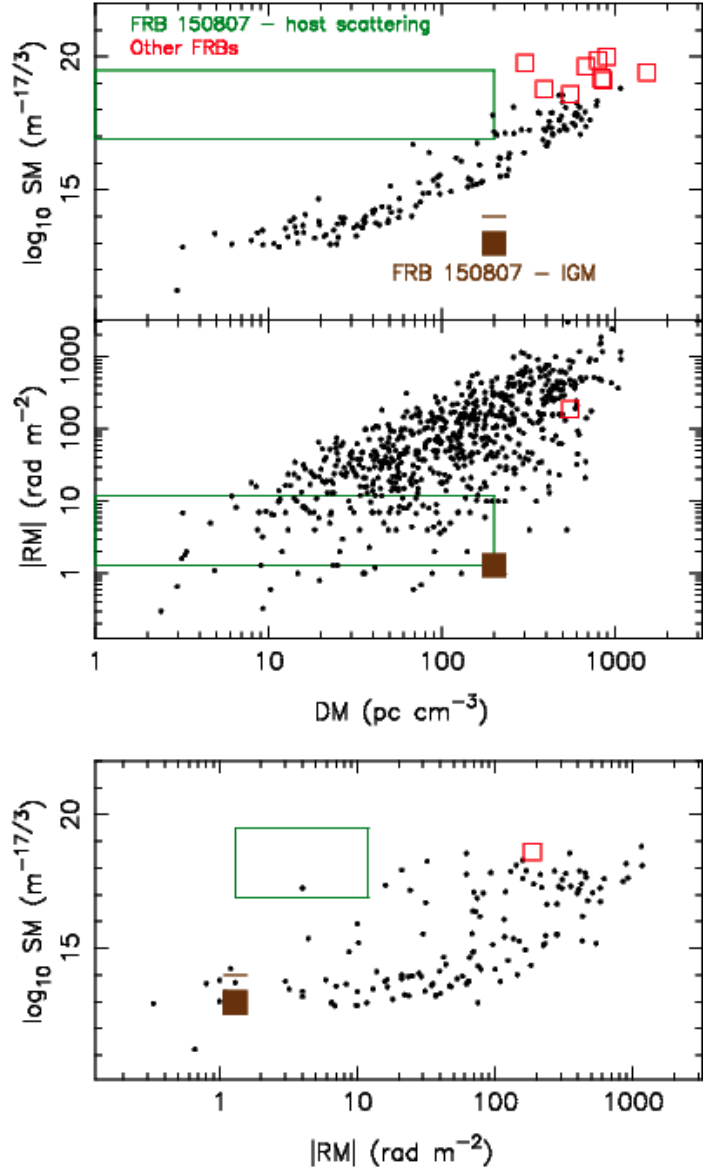

**Fig. 3. Properties of the plasma toward FRBs and pulsars.** *Green boxes:* possible ranges of DMs, RMs and SMs contributed by a putative host galaxy of FRB 150807, assuming negligible RM and SM contributions from the IGM. The upper limit of the FRB 150807 SM range assumes a screen at 0.1 kpc from the source, the source at $D = 100$ Mpc, and the upper limit (27 μs) on $\tau_d$, and the lower limit assumes a screen at 10 kpc, the source at $D = 500$ Mpc, and $\tau_d = 1.6$ μs. The DM range for FRB 150807 is bounded by the total estimated extragalactic component. The RM range for FRB 150807 is bounded by its total detected RM, and by its extragalactic value corrected for the Milky-Way contribution. *Brown squares:* The DM, RM and SM of FRB 150807 assuming contributions predominantly from the IGM, and corrected for Milky Way contributions. The brown bars were calculated assuming $\tau_d = 27$ μs. *Red squares:* DMs, RMs, and SMs of other FRBs (*31*). We assumed extragalactic DMs contributed entirely by putative host galaxies, and converted the scattering measurements into SMs assuming screens at 10 kpc from the sources and $D = 100$ Mpc. *Black dots:* DMs, RMs and SMs of Milky-Way pulsars (*32*).

The SMs were calculated assuming intervening scattering screens equidistant from the Earth and the burst source.

| Time of burst peak | UT (Ref. freq. 1207 MHz) | 2015 Aug 07 17:53:55.77991±0.00009 |
|---|---|---|
| Position (J2000) | Right Ascension<br>Declination<br>Galactic longitude ($l$)<br>Galactic latitude ($b$) | 22:40:23±4<br>-53:16±2<br>336.71±0.03°<br>-54.40+0.03° |
| Peak flux (averaged over 1182-1519.5 MHz) | measured<br>implied (boresight)* | 12.2±0.1 Jy<br>120±30 Jy |
| Peak flux in dynamic spectrum | measured<br>implied (boresight)* | 128±5 Jy<br>1000±300 Jy |
| Fluence | measured<br>implied (boresight)* | 4.6±4 Jy ms<br>50±20 Jy ms |
| Spectral index $\alpha$, for assumed spectral form proportional to $f^\alpha$ | measured<br>implied (boresight)* | -7±2<br>-5±2 |
| Width | At half maximum amplitude | 0.35±0.05 ms |
| Dispersion measure (DM) | total<br>extra-Galactic† | 266.5±0.1 pc cm$^{-3}$<br>200±20 pc cm$^{-3}$ |
| Faraday rotation (RM) | total<br>extra-Galactic‡ | 12.0±7 rad m$^{-2}$<br>-1.3±7 rad m$^{-2}$ |
| Mean line-of-sight parallel magnetic field, $B_\parallel$ | total<br>extra-Galactic†,‡ (2σ limit) | 56±5 nG<br>< 21 nG |
| Inferred scintillation bandwidth | half-width at e$^{-1}$ of the spectrum-autocorrelation peak | 100±50 kHz |

**Table 1. Properties of FRB 150807.** *After correcting for the maximum-likelihood beam attenuation. †After removing the Milky Way DM contribution, estimated to be 70±10 pc cm$^{-3}$. ‡ After removing the RM of the nearby pulsar PSR J2241−5236, adopted as the Milky Way RM contribution, and assuming a |RM| of 2.7 rad m$^{-2}$ and an extragalactic DM of 160 pc cm$^{-3}$.

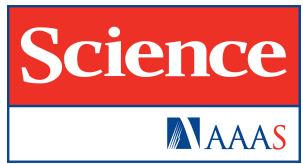

Supplementary Materials for

The magnetic field and turbulence of the cosmic web measured using a brilliant fast radio burst

V. Ravi, R. M. Shannon, M. Bailes, K. Bannister, S. Bhandari, N. D. R. Bhat, S. Burke-Spolaor, M. Caleb, C. Flynn, A. Jameson, S. Johnston, E. F. Keane, M. Kerr, C. Tiburzi, A. V. Tuntsov, H. K. Vedantham

Correspondence to: vikram@caltech.edu, Ryan.Shannon@csiro.au

**This PDF file includes:**

    Materials and Methods
    Supplementary Text
    Figs. S1 to S11
    Tables S1 to S4

**Materials and Methods**

1 – Observing set-up for the detection of FRB 150807

We detected the fast radio burst (FRB 150807) in real time with the 64-m Parkes Radio Telescope, using the 21-cm Multibeam receiver (MB21) receiver (*6*) and the Berkeley Parkes Swinburne Recorder backend (BPSR, Ref. *7*). The MB21 system comprises 13 circular stepped-feed horns arranged in a hexagonal pattern in the prime focal plane of the dish. Inner and outer rings of six horns surround a central horn, with all equipped with two linearly polarized receivers in orthogonal orientations. Our adopted model for the far-field radiation pattern of this 13-beam system, discussed below, is shown in Fig. S1 (Ref. *38*).

The receiver and down-conversion system in our observations provide 400 MHz of bandwidth between 1182 and 1582 MHz. The frequency band above 1519.5 MHz is unusable because of radio-frequency interference (RFI) from the *Thuraya-3* geostationary satellite, and is filtered out prior to digitization of the voltage signal. The real-time system (*8, 9*) allowed us to collect full-polarization data around candidate bursts.

Our observations were undertaken during the Parkes All Radio Transients in the skY (PARTY) survey (project P883), which intended to search for FRBs in the lower frequency *50cm* band (700-764 MHz). After the commencement of the project, a 4G telephone base station was deployed in nearby Alectown, New South Wales, introducing strong RFI in this band, resulting in us altering our observing plan and observing band. We instead tracked the high Galactic latitude millisecond pulsar PSR J2241−5236 with the central beam of the MB21 to obtain timing data for pulsar timing array projects (*39, 40*), and simultaneously searched for FRBs with all beams of the system. As is standard practice for precision timing observations at Parkes, the MB21 beams were held at a fixed position angle relative to the dish; consequently, the outer beams rotated on the sky over an observation because Parkes is an altitude-azimuth mounted telescope. FRB 150807 was detected in an observation that commenced at 17:07:28 UT on 2015 August 7th, and the *Heimdall* real-time burst search pipeline identified FRB 150807 2787.83 seconds into the observation in beams 5 and 12, with reported signal-to-noise ratios of 50 and 8 respectively. The system automatically stored 1.024 s of full-polarization (4-product) data for all beams of the MB21, temporally centered on the FRB. These data comprised of a time series for each polarization of 1024-channel, 8-bit spectra with 0.390625 MHz channel widths, integrated for 64 microseconds. Inspection of the data recorded for all other beams revealed no detections with signal to noise ratios (SNRs) greater than 3. At the time the FRB was detected, beam 5 was centered at coordinates RA 22:41:33.75, DEC −53:05:39.9. The pointing accuracy of the Parkes telescope is approximately 11 arcsec. The azimuth and altitude of the telescope were 220.0603 and 59.8699 deg. respectively when the FRB was detected.

2 – Polarization and flux calibration

We applied standard techniques to calibrate the instrumental gain and polarization of the MB21. To calibrate the polarization of the FRB, we performed a two-minute scan of a modulated broadband noise source injected into all feed horns of the MB21 at a 45-degree angle to the orthogonal linear probes. This scan was conducted approximately 2.5 hr after the FRB was detected. Using the BPSR backend, the data were folded in real time and integrated into 1024 bins across the period of the modulated noise source, preserving all polarization information. At this time, the central beam of the MB21 was still pointed at the celestial coordinates of PSR J2241−5236. We rotated the angle of the receiver so the outer beams were observing the same position in the sky as when the FRB was detected. The flux scale for beams 5 and 12 was determined using standard flux-calibration observations 20 days after the FRB was detected by comparing the system response to the noise diode, and then to that of the radio galaxy Hydra A. The MB21 system shows very little gain variation (*39*) and we estimate the uncertainty in the flux scale to be less than 1%.

We excised data containing persistent narrow-band RFI and removed the channels that were pre-filtered for the *Thuraya-3* satellite. The header metadata contained in raw BPSR files were found to be incorrect, with the receiver symmetry angle (90 deg.) and backend phase convention (-1) parameters incorrectly valued. This resulted in incorrect polarimetry of previous Parkes FRBs. We then used the PSRCHIVE (*41, 42*) programs *fluxcal* and *pac* to calibrate the data. The specific calibration algorithm used was 'SingleAxis+Flux' (*43*) that assumes the feed probes are orthogonal, linearly polarized receptors with no cross coupling. However, the flux calibrator observations enable the relative illuminations of the receptors by the polarized noise source to be measured. In summary, this calibration process corrects for the absolute frequency dependent instrumental gain, and the instrumental response to a polarized source.

Through extensive testing, it is known that cross coupling at the 10% level (in power) exists between the receptors in the central beam of the MB21, which requires long tracks of polarized sources such as pulsars to calibrate (*44*). The discrepancy between these measurements and the MB21 design has not been resolved and detailed studies of other beams of the MB21 have not been carried out. However, extrapolating from the central beam, we expect that the fractional error in the measurement of the polarization properties of FRB 150807 is less than 10%. We confirmed this level of reliability (and the robustness of our polarization measurement) by observing pulsars with known polarization properties at many positions relative to the boresight of beam 5, and at the modeled position (relative to beam center) of the FRB.

3 – Localization of FRB 150807, and correction for the off-axis attenuation

The highly significant detections of FRB 150807 in both beams 5 and 12 and non-detections in other beams allowed us to constrain the FRB position to a region much smaller than the half-power beam width of any individual beam. It also enabled us to correct for the frequency-dependent attenuation of the measured FRB flux density due to beam attenuation.

We computed an analytic model for the radiation pattern of the MB21 based on expressions for off-axis feeds (*45*), and published details of the geometry of the dish and the MB21 (*6, 46*). We assumed a 5-m diameter blockage from the prime focus cabin, and a Gaussian taper of the radiation pattern of each beam with a spillover efficiency of 96% at the middle of the band. We also assumed that the offset distance from the prime focus is a negligible fraction of the focal length, implying that each beam is effectively tilted toward the vertex of the dish. We additionally treated each feed in isolation, not accounting for the presence of other feeds in the MB21 array when calculating the radiation pattern of a given feed. The model was computed over a 1000×1000 point grid spanning 3×3 deg. centered on the central beam of the MB21. The code used to generate the radiation pattern is provided as supplementary files (gen_maps.py, comalobes.py).

The model radiation pattern, shown in Fig. S1, at a frequency of 1357 MHz, was tested against observations of the bright pulsar PSR J1644−4559 at 100 positions indicated in the top-left panel of Fig. S1. This pulsar was chosen because it has both a large flux density (0.31 Jy averaged over the pulse period) and a large dispersion measure [DM = 478.8 $cm^{-3}$ pc, (*47*)]. The pulsar is also strongly scattered and so does not show significant temporal or frequency variation associated with diffractive scintillation. At each pointing, the data for each beam were folded according to the pulsar ephemeris by BPSR over a two-minute interval. For the beams of interest (we show results for beams 2 and 9 in Fig. S2), we calculated the SNR of the folded total-intensity pulse profile. For each observation, we use the matched filter SNR, calculated using an analytic model for the pulse profile.

In Fig. S2, we show a comparison between the predicted and measured SNRs for beams 2 and 9, scaled to those at the pointing with the highest measured SNR for each beam. Measurements are shown for each of the scans labeled in Fig. S1, where the first pointing is the one closest to the numerical label. The predictions and measurements both correspond to the average within a 50 MHz band centered on 1357 MHz.

The match between the predictions and measurements is generally good, with discrepancies of less than a factor of two evident for the pointings where the response is >1% of the maximum for each beam. We include this uncertainty in our localization of the FRB. Although the model systematically over-predicts the response for the pointings most distant from beam centers, these pointings are not relevant to our analysis. This is likely caused by the assumption of negligible feed offsets in the focal plane relative to the dish focal length.

To localize FRB 150807, we measured the SNRs of the flux-calibrated FRB 150807 detections in beams 5 and 12 of the MB21 in sub-bands of 50 MHz bandwidth centered on 1207, 1257, 1307 and 1357 MHz, using our models for the M21 radiation patterns in these bands. In beam 12, the FRB was not detected at higher frequencies and is observed to have a more steeply negative spectrum. This is consistent with greater beam attenuation at higher frequencies (where the beam size is smaller), implying an FRB position closer to the first null of the main lobe. We created Monte-Carlo realizations of each observation based on the measurement errors to calculate a containment region for the burst. For each realization, we then found the points in the beam model that were within a factor of 4 of the measured ratio between the beam 5 and beam 12 SNRs, in all frequency bands. We applied the additional constraint that the FRB was not detected with an SNR ≥ 3 in any other beam. Thus, for each realization, we produced single-bit maps with values of 1 if the FRB was consistent with the position and 0 if it was not. We performed 1000 realizations, and summed all these maps. The resulting 95% containment region for the FRB position, converted to sky coordinates for the time the FRB was detected, is shown in Fig. 2 of the main text. The most likely position of the burst is shown in Fig. S1. We also include an additional 11-arcsec. uncertainty to account for the pointing uncertainty of the Parkes telescope. The localization results also enabled us to derive frequency-dependent corrections to the FRB flux density from the MB21 radiation pattern model, shown in Fig. S3, that we applied to the calibrated FRB data.

Observations of pulsars at different positions within the detected beams confirmed the reliability of our measurement of linear polarization and Faraday rotation (see below). In Fig. S4, we show integrated pulse profiles in the Stokes Q, U and V parameters for the bright pulsar PSR J1644–4559 obtained on-axis for beam 5 of the MB21 at the most likely off-axis position of the burst. Although stochastic pulse-to-pulse variations intrinsic to the pulsar add some uncertainty to the measurements in the short 2-min observations, we find that the recovered polarization properties are consistent with published polarimetry of the pulsar (*48*). We also estimated the Faraday rotation measure of the pulsar using the PSRCHIVE program *rmfit* for both observations. For the on-axis observation, we obtained RM = -621±2 rad m$^{-2}$, and for the off-axis observation we obtained RM = -620±20 rad m$^{-2}$. In addition to these values being consistent with each other, they are consistent with the previously measured value of RM = -617±1 rad m$^{-2}$ (*49*).

4 – Faraday rotation measure estimation for FRB 150807

We assume that FRB 150807 represents emission from a single point source and is the only pulsed source of polarized emission along the sightline at its DM when we detected it. The point-source assumption is justified because causality arguments limit the sizes of FRB emission regions to $< c\Delta t \sim 300$ km, where *c* is the vacuum speed of light and $\Delta t$ is the FRB timescale (*50*). Furthermore, from the burst dynamic spectrum, we infer only modest levels of scattering of the burst during its propagation. Substantially different propagation paths through a dense, magnetized medium could result in the burst emission being consistent with multiple rotation measures (RMs) (*51*), but would likely depolarize the burst unless the burst resulted from separate multiple images. As we subtract the mean off-pulse emission in all Stokes parameters prior to our analysis, other polarized emission does not affect RM estimation of the pulse.

The small inferred angular size of FRB 150807 and the modest levels of multi-path propagation imply that an RM synthesis analysis will result in the detection of emission at a single Faraday depth (*52*). In this case, Faraday depth and RM are equivalent. Further, the high level of detected linear polarization in the average pulse profile (Fig. 1 B) implies that the level of Faraday rotation is small within the bandwidths of the original 1024 channels. Prior to analysis, we therefore (a) averaged the data to 32 frequency channels to smooth over the fine frequency-structure in the burst intensity and increase the SNR of our polarization measurements, (b) derived the Stokes Q and U spectra of the burst by performing a signal-to-noise weighted average over the pulse profile and subtracting the off-pulse levels, and (c) normalized the Stokes Q and U spectra by $(Q^2+U^2)^{1/2}$ to correct for the spectral variation of the burst.

We measured the RM of FRB 150807 using multiple techniques, which gave consistent results. In the first technique, we derived the RM from the variation of the position angle of the linear polarization vector across the band. We did this both with a least-square fitting algorithm and with the program *rmfit* that is part of PSRCHIVE. The result was RM = 11.50±8 rad m$^{-2}$.

In the second, we used Bayesian methodology to find the posterior distribution of the RM given the data as the probability of obtaining the RM-corrected Stokes Q and U data. The Bayesian methodology enabled us to explore the range of uncertainty in the RM, by deriving its posterior distribution given the data. The uncertainty in each frequency channel was derived from the variance of the off-pulse data. We assumed a uniform (flat) prior in the RM in the range -4000 rad m$^{-2}$ to 4000 rad m$^{-2}$, used a Markov Chain Monte Carlo algorithm to sample the posterior RM distribution, and determined the burst-averaged RM to be 12.9±7 rad m$^{-2}$. The data used in our analysis are shown in Fig. S5, along with the model fit. We adopt this measurement for the RM, as it does not suffer from the possibility of underestimated position-angle measurement errors.

Finally, we considered the implications of relaxing our assumption of the burst being consistent with emission at a single Faraday depth, by performing a full RM synthesis analysis (*52*). We averaged the data to 256 spectral channels in order to smooth over rapid fluctuations in the spectrum, and did not normalize the Stokes Q and U spectra. Although not normalizing the spectra can bias the RM estimation because of intrinsic spectral variations, the almost 100% linear polarization fraction of the burst means that intrinsic spectral variations are degenerate with emission at different Faraday depths. The results of our RM synthesis are shown in Fig. S6. The first peak is at a Faraday depth consistent with our previous RM measurements. The second peak, at a Faraday depth of approx. 180 rad m$^{-2}$, could be caused by the broad spectral variations in the total-intensity spectrum of the burst, due to the Faraday depolarization effect. However, the compact nature of the burst progenitor, inferred from the detected burst timescale, makes it extremely unlikely that emission is present at multiple Faraday depths. The 180 rad m$^{-2}$ feature is hence likely to be a systematic artifact, associated with bandpass of the multibeam system.

The expected RM along this line of sight contributed by the ionosphere at the time of the burst, which we calculated using the publicly available ionFR software package (*53*), is 0.9 rad m$^{-2}$. The Galactic foreground RM along this sightline is poorly determined in the literature owing to sparse measurements of the RMs of extragalactic sources; the 95% confidence interval for the foreground RM is between −20.7 rad m$^{-2}$ and 23.0 rad m$^{-2}$ (*25*). However, the millisecond pulsar PSR J2241−5236, which is 32 arcmin distant from the FRB and has a DM of 11.41±1 cm$^{-3}$ pc, has an RM of 13.3±1 rad m$^{-2}$ (*22*). The DM of this pulsar is consistent with a location at approximately half the scale height of thick disk of warm ionized gas (*12-14*), and RMs induced by magnetic fields in the Galactic halo at high Galactic latitudes are expected to be substantially smaller than the pulsar RM (*15*). We therefore adopt the pulsar RM as the Galactic contribution along this line of sight. We note that contribution to the extragalactic RM from the Local Group of galaxies is only expected to be a few rad m$^{-2}$ (*54*).

5 – Follow-up observations

We conducted follow-up observations of the field of FRB 150807 soon after the burst epoch with the Australia Telescope Compact Array [ATCA, (*55*)] radio interferometer and the *Swift* space-borne observatory. As we had not at the time obtained the accurate localization that we describe in this paper, we instead surveyed a 0.5×0.5 deg. region centered on the location of beam 5 when the burst was detected. This included the 95% containment region we later identified for the burst.

*5.1 – ATCA observations*

The ATCA observations were conducted on four epochs between 1 day and 1 month after FRB 150807. Details of the observations are given in Table S1. We obtained visibility data (averaged into 10 second integrations) in two 2.048 GHz bands, each split into 2048 channels, centered on 5.5 GHz and 7.5 GHz. To image the 0.5×0.5 deg. field it was necessary to use 42 mosaicked pointings, spaced by the half-power beam-width of the primary beam of the 22-metre dish antennas at 8.5 GHz. The flux density scale and instrumental bandpass were calibrated using the primary ATCA calibrator, radio galaxy PKS B1934−638. We also observed an unresolved quasar PKS B2204−540 located near the target field every 30 min as a secondary phase calibrator to calibrate the time-varying instrumental and atmospheric gains and phases.

The ATCA data were reduced with the *Miriad* software package (*56*), following standard processing. We produced multi-frequency synthesis images of the five pointings closest to the containment region for the burst, with 'robust' weighting (robust parameter 0.5) to optimize point-source sensitivity. The array was in very compact configurations at all epochs (see Table S1), with 5 of the 6 antennas within 214 meters of one another, and the sixth approximately 3 km away. As a result, useful images could only be made with the five closely spaced antennas during these epochs, with the spatial resolution of our images being at best 60 arcsec. (Table S1).

No sources were detected within the containment region for FRB 150807 at any epoch and frequency band. Typical total-intensity rms noise values within the containment region were 0.1 mJy, which are consistent with a combination of thermal noise, source confusion, and the containment region lying at the edge of the 0.5×0.5 deg. field. 3σ upper limits on source flux densities are given in Table S1. The sensitivity of our observations was approximately a factor of 8 worse than previous FRB follow-up observations (*57*) that revealed a candidate FRB radio afterglow.

*5.2 – Swift observations*

We conducted four pointings with the *Swift* X-ray Telescope [XRT, (*58*)] and co-aligned UV-Optical Telescope [UVOT, (*59*)] to tile most of the 0.5×0.5 deg. field. The pointing that included the FRB containment region was conducted at UT 2015 August 08 10:57 (17 hours after FRB 150807) and lasted for 930 seconds, with the XRT in proportional-counting mode. Unfortunately, the reduced UVOT field of view with respect to the 23.6×23.6 arcmin XRT field of view meant that the UVOT observations did not cover the containment region. The Swift Burst Alert Telescope issued no alerts during this time.

We processed the data using the standard script (*xrtpipeline*) distributed with HEASOFT (*60*), and the latest *Swift* XRT calibrations. No sources were detected within the entire XRT field of view for the pointing of interest. The limiting sensitivity in 930 seconds for the XRT is $6.6\times10^{-14}$ erg cm$^{-2}$ s$^{-1}$ in the 0.2-10 keV band.

*5.3 – High time resolution observations at Parkes and Molonglo*

To date, we have observed the location of FRB 150807 with the Parkes telescope for 90 hr, and the upgraded Molonglo Observatory Synthesis Telescope (UTMOST) for 125 hr, to search for repeat bursts. No bursts at the same DM as FRB 150807 were detected with widths <~30 ms. The Parkes observations were undertaken with either the central beam of the MB21 receiver using an identical configuration to the discovery observations, or with the single-pixel H-OH receiver in the same band and with similar sensitivity, and were mostly pointed directly at the correct localization. The limiting (10σ) fluence for a repeat burst was 0.3 Jy ms, assuming a width given by the dispersion smearing timescale. The UTMOST observations were conducted in the standard FRB search configuration (*61*) at 843 MHz, but with somewhat improved sensitivity than previous searches. Although the telescope was pointed at the millisecond pulsar PSR J2241–5236, the wide sky-response of UTMOST meant that the sensitivity to repeat bursts from FRB 150807 was only degraded by a factor of 2. The limiting (10σ) fluence for a repeat burst was therefore 38 Jy ms, assuming a width given by the dispersion smearing timescale.

**Supplementary Text**

S1 – Strong diffractive scintillation in the dynamic spectrum of FRB 150807

*S1.1 – Exponential intensity variations in the dynamic spectrum*

In this section, we show that the strong spectral modulation of FRB 150807, which is indicative of microsecond-scale structure in the temporal profile, has an exponential intensity distribution. An analysis of pulses from bright pulsars confirms that instrumental effects do not cause the observed modulation.

The width of FRB 150807 is consistent with a temporally unresolved pulse smeared by uncorrected dispersion within individual spectrometer channels. In Fig. S7, we show measurements of the pulse width at different frequencies, assuming a Gaussian profile. The Gaussian profile approximately corresponds to the expected response given the 2-tap polyphase filterbank design of the BPSR spectrometer, with the pulse width evolving as frequency to the power -3.0±0.4. The measured pulse widths are marginally above the expected smearing timescales within individual channels at different frequencies, assuming no overlap in the channel frequency-responses. A more rigorous comparison would involve careful modeling of the frequency responses of individual channels of the BPSR polyphase filterbank design.

Despite the unresolved nature of the burst, the dedispersed dynamic spectrum of FRB 150807, displayed in Fig. 1 C of the main text, shows evidence for decorrelation in both frequency and time. That is, structures appear to be present in specific time-frequency bins that are confined to narrow ranges in time and frequency. The timescales of the structures appear substantially less than the frequency-averaged burst temporal width. We use two complementary approaches to measure the characteristic scale of the structure in the dynamic spectrum, using both the spectral and temporal modulation.

Direct estimation of the bandwidth using the spectral modulation alone is difficult because the pattern is marginally unresolved relative to the spectral resolution of BPSR. From the spectral modulation we estimate the characteristic decorrelation bandwidth using the frequency-autocovariance function (Fig. S8 A), defined to be the lag at which the autocovariance decays to $1/e$ of its power, to be 200±100 kHz. This scale is less than the BPSR spectrometer channel resolution of 390.625 kHz, and is likely biased upwards by receiver noise.

The characteristic decorrelation timescale of the dynamic spectrum is better resolved than the frequency-decorrelation scale. From the time-autocovariance function of the burst (Fig. S8 B), we measure a characteristic decorrelation timescale of 100±50 μs.

We consider two possible interpretations for the frequency- and time-decorrelations evident in the autocovariance functions: (a) that they represent thermal noise (which can include self noise from the source); and (b) that the noise is the manifestation of temporal structure in the pulse (*28*), which can be caused by scattering in inhomogeneous plasma (*18*), or is intrinsic to the emission mechanism (*62*). In case (a), the measured radio power in a given time-frequency bin with a time $t_i$ and frequency $f_j$, following subtraction of the off-burst baseline, is given by

$$P_{ij} = S_j F_i + N[0,(SEFD + S_j F_i)(\Delta f \Delta t)^{-1/2}] \qquad (S1)$$

where $S_j$ is the intrinsic burst spectrum, $F_i$ is the frequency-averaged temporal profile of the burst normalized by the peak, *SEFD* is the system-equivalent flux-density, $\Delta f$ is the frequency bin-width, and $\Delta t$ is the time bin-width. $N(\mu,\sigma)$ represents a normal random variable distributed with mean $\mu$ and standard deviation $\sigma$. We assume negligible systematic correlation between adjacent time and frequency bins. In estimating $S_j$, we do not correct for the beam attenuation of the burst flux density. This is because, in case (a), the total system temperature is a combination of a systematic component (including the intrinsic receiver temperature, spill-over radiation, and RFI), a sky-background component (which is negligible for our high Galactic latitude pointing), and a source component. The source component is determined by multiplying the source flux density by the antenna gain, which is much lower for the FRB than for an on-axis source. We adopt an *SEFD* of 40 Jy, including all aforementioned components besides the source.

If the burst is composed of microsecond-scale structures (case b), the flux distribution will be modified (*28*). $P_{ij}$ will be distributed about the mean $S_iF_j$ approximately as an exponentially modified Gaussian distribution (the convolution of thermal noise with an exponential distribution), with a rate parameter that is determined by the properties of the microsecond-scale structures.

We distinguish cases (a) and (b) by analyzing the distribution of measurements of $P_{ij}$ in the dynamic spectrum of FRB 150807. The dynamic-spectral measurements are preferred in this analysis over the time-averaged channel intensities because of temporal variability across the pulse. In case (b), the combination of the unresolved pulse and spectral peaks with bandwidths less than the spectrometer channel width would suggest that the time-decorrelation is an artifact of the incoherent dedispersion process. Fixed time-delays are added to all measurements in a single channel to correct for dispersion, implying that the spectral peaks within a single channel will be incorrectly delayed (and hence spread in time) unless they all lie exactly at the center-frequency of the channel. Our effective frequency resolution is therefore improved to the bandwidth corresponding to the dispersion smearing timescale of a single 64-microsecond time-sample, which is 76.3 kHz.

Analysis of the dynamic spectrum rules out case (a). In doing so, we estimate $F_i$ and $S_j$ at the native BPSR time resolution of 64 μs and the native BPSR frequency resolution of 390.625 kHz respectively. We first subtract a mean baseline level for each channel estimated from the off-pulse data. To estimate $F_i$, we average the dynamic spectrum across all frequency channels unaffected by RFI and normalize by the peak value. To estimate $S_j$, we average the data in time within all time-bins of $F_i$ above the noise. In order to achieve better SNR in estimating $S_j$, which is the intrinsic burst spectrum, we smooth the measured $S_j$ with a top hat function of width 20 channels. Trials of top hat widths between 10 and 50 channels did not significantly alter our final results. We then subtract our estimated $S_iF_j$ from the measurements of $P_{ij}$, and divide by the expected standard deviation in Equation (S1) assuming an SEFD of 40 Jy. If case (a) were true, the resulting data should be distributed as $N(0,1)$. We show a histogram of these data within the 14 significant time-bins of FRB 150807, and between 1182 – 1382 MHz where the burst is primarily detected, in Figure S9. We also show a histogram of data processed in exactly the same way, within the same frequency range, but for 14 time-bins adjacent to the FRB.

The on-pulse data are clearly inconsistent with a normal distribution with zero mean and unit variance. Instead, they appear consistent with an exponentially modified normal distribution with a normal component given by $N(-2.0,1)$ and a rate parameter of 0.57, as shown in the Figure. This suggests that case (b) is preferred over case (a). That is, the intensity variations in the dynamic spectrum are consistent with the sum of a normal random variable consistent with the estimated radiometer noise level, and an exponentially distributed component expected given microsecond-scale structure in the pulse. The mean of the distribution is slightly different from expectation, because our estimator for the mean will be biased for exponentially distributed data.

On the other hand, the off-pulse data are consistent with the $N(0,1)$ distribution, as expected for thermal noise with the predicted power. Further, we verified that the exponential statistics observed on-pulse are not a systematic effect by analysing individual pulses, with comparable SNRs to FRB 150807, from the Vela pulsar (PSR J0835−4510) in the same way. The observations were conducted with exactly the same receiving system. Exponentially distributed intensities are not expected in the dynamic spectra of Vela pulses, and were indeed not observed.

We therefore conclude that the time- and frequency-decorrelation of the dynamic spectrum of FRB 150807 is inconsistent with thermal noise. In this case, the decorrelation timescale of the dynamic spectrum will correspond to the characteristic dispersion smearing timescale of the frequency-peaks (and not the channel bandwidth).

We can thus better estimate the bandwidth of the spectral peaks from the temporal decorrelation of the dynamic spectrum, rather than from the spectral decorrelation. A timescale of 100 μs corresponds to a bandwidth of 100 kHz at 1.3 GHz, somewhat smaller than the decorrelation bandwidth of 200 kHz ±100 kHz inferred directly from the spectrum. This discrepancy is likely caused by an upward bias to the decorrelation bandwidth measurement introduced by noise. In this paper, we therefore adopt a decorrelation bandwidth of 100 kHz, with a measurement uncertainty of 50 kHz.

*S1.2 – What causes the microstructure in FRB 150807?*

There are two possible sources for the micro-second/kilohertz structure in the dynamic spectrum of FRB 150807. The structure could be intrinsic to the emission mechanism or the result of propagation through inhomogeneous plasma. In the latter case, the variations in the burst spectrum would be interpreted as strong diffractive scintillation (*18*).

If the structure in the burst dynamic spectrum were scintillation, we would expect the decorrelation bandwidth to scale proportionally to $f^{-4}$. However, the resolution of our instrument combined with the low burst intensity in the upper part of our band is insufficient to ascertain whether this scaling is present for the 100 kHz structure. Without this information, it is difficult to conclusively distinguish between scattering and intrinsic pulse structure on ~10 μs scales [1/(100 kHz)] as causing the 100 kHz structure (*28*). A combined analysis of the intrinsic burst profile and propagation effects as beyond the scope of this paper. Such an analysis would rely on some knowledge of the scattering geometry and statistics of the pulse broadening function, which we lack. We would also have to understand the response of the BPSR spectrometer to a signal that is significantly non-stationary over the timescale of the spectrum estimation, which would bias the Fourier representation of the signal. However if the intrinsic pulse structure interpretation is correct, it is surprising that spectral intensity modulations are not evident in other FRBs, in particular those detected with high SNRs such as FRB 110220 (*2*) and FRB 150418 (*57*).

*S1.3 – Summary of scattering theory*

Strong diffractive scintillations (intensity modulation index of order unity) are caused by significant phase variations at the observer on a scale less than the Fresnel scale, $(L/cv)^{1/2}$, where $L$ is the characteristic distance to the scattering medium, and $v$ is the radio frequency. Such phase variations, manifested as strong scintillation of compact sources such as pulsars and AGN, are ascribed to scattering caused by density (and hence refractive-index) fluctuations in the free-electron distribution of the Milky Way. However, for FRBs, which propagate over extragalactic distances and possibly through substantial plasma columns in host galaxies, the effects of plasma density fluctuations in the intergalactic medium (IGM) and the possible host galaxies must also be considered.

Relationships between observable effects of radio wave propagation in the medium and the properties of the media themselves are also well established (*18*). To a good approximation (15%) under the conditions we consider in this paper, the characteristic delay of scattered rays, $\tau_d$, can be related to the scintillation bandwidth, $f_d$, as $\tau_d = 1/(2\pi f_d)$. Scattering also causes point sources to appear broadened in angular extent, and is characterized using a characteristic broadening angle, $\theta$, which is the angular extent over which the observed visibility function of the source is reduced by $e^{-1}$. A ray-optics analysis is used to relate $\theta$ to the scattering measure (SM), which is a widely used physical quantity characterizing the amplitude of the density fluctuations in the line-of-sight free-electron distribution. The relationship between $\tau_d$, $\theta$, and SM depends on the assumed geometry of the scattering medium. In the main text, we consider two commonly applied geometries in this paper: a `thin screen' model where all scattering is confined to a single distance along the line of sight, and a `uniform slab' model where the properties of the scattering medium are homogeneous along the entire line of sight. The expressions for the SM in each case in the main text assume a Kolmogorov turbulence spectrum for the density fluctuations, which is commonly observed in the interstellar medium, and also assume that the characteristic fluctuation scale, $c/(2^{1/2}\pi\theta f)$, is not near either the inner or outer scale of the fluctuation spectrum. For constant SM, the expressions imply $\tau_d \sim f^{-4.4}$.

*S1.4 – Upper limit on the scatter broadening timescale*

If, alternatively, the spectral variation in FRB 150807 is caused by intrinsic structure on ~10 microsecond scales, which could cause the observed exponential intensity variations in the dynamic spectrum, we can place an upper limit on the scatter broadening timescale, $\tau_d$, directly from the pulse width. We consider a model for the pulse where a one-sided exponential with characteristic scale $\tau_d$, assumed to be proportional to $\sim f^{-4.4}$, is convolved with the dispersion smearing profile. We explore the posterior distributions of the model parameters given the data using a technique identical to that used by Ravi, Shannon & Jameson (*63*) to model FRB 131104 (their Model 2 with fixed $\alpha$=4.4). We find that $\tau_d$ is consistent with 0, and place a 95% confidence upper bound on $\tau_d$ at 1.3 GHz of 27 μs. Our estimate of the marginalized posterior density of $\tau_d$ is shown in Fig. S10.

S2 – Constraints on the source of FRB 150807 using archival near-infrared data

The properties of the host system and scattering region can be constrained by cross-matching our constraint on the FRB position with archival observations of the field. The best archival observations for this purpose are from the Visible and Infrared Survey Telescope for Astronomy (VISTA) Hemisphere Survey (VHS), which utilizes a 4-metre class telescope to image the southern sky in six different wavelength bands (*11, 64, 65*). We used observations from Data Release 3 (*65*).

We obtained 20×20 arcmin VHS image stacks in the J, H and Ks bands, centered on the 95% containment region of the FRB. The J and Ks band data were the most sensitive. We used the *SExtractor* source extraction routine (*66*) with standard parameters to find and estimate source magnitudes, and perform morphological star-galaxy classification. The 4σ limiting magnitudes in the J and Ks bands were respectively 19.7 and 19.5. A color-magnitude diagram for the sources within the containment region is shown in Fig. S11.

Only nine sources were detected at >4σ. Their measured parameters are given in Table S2, along with cross-identifications from the Widefield Infrared Survey Explorer (WISE) mission (*67*). The bright galaxy VHS7 is also catalogued in the Muenster Red Sky Survey as MRSS 190-051545, with a R-band magnitude of 18.8. Four of the sources have cross-identifications in the SuperCOSMOS catalog (*68*), as listed in Table S3. No corrections for Galactic extinction have been applied to the data in Table S3, but galactic extinction is likely to be <0.02 magnitudes along the high galactic latitude line-of-sight (b ~ 55 deg.). The sources VHS3, VHS8 and VHS9 are likely to be main-sequence stars in the Milky Way, as indicated by their moderately red optical colors, brightnesses and *SExtractor* classifications. They are all also listed in the USNO-B1.0 star catalog (*69*).

We estimated the redshift and galaxy type of VHS7 using two publicly available software packages. The first, *hyperz* (*70*), provided a best-fit value corresponding to a S0 galaxy of $z$ = 0.37, with a 90% confidence interval lying between 0.24 and 0.42. The second, *eazy* (*71*), also suggested a 'massive red galaxy' at $z$ = 0.32, with a 95% confidence interval of 0.20 to 0.36. As the *eazy* results also include the WISE magnitudes, we adopt these. Given the Planck (*72*) cosmological parameters, we infer a luminosity distance of 1.7 Gpc for VHS7, with a 95% confidence interval of between 1.0 Gpc and 2.0 Gpc.

The five VHS sources that cannot clearly be classified as stars or galaxies have sufficiently red J-Ks color indices as to suggest that they are galaxies (*73*). Two independent simulations of the Milky Way stellar content of a 3 deg$^2$ region of the sky at the Galactic coordinates of FRB 150807 using the Besancon stellar population synthesis model (*74*) revealed no stars fainter than $J = 18.1$. Therefore, it is highly likely that the five unidentified VHS sources are galaxies. As they are all approximately two magnitudes fainter than VHS7, but have consistent colors, it is likely that they are also more distant (we do not have sufficient photometric information to estimate their redshifts). Hence, we identify VHS7 as the closest candidate counterpart to FRB 150807 among the sources detected in the VHS data.

If the five VHS sources with unclear classifications are galaxies, we can quantify our expectation for their distances using simulations of the galaxy content of the Universe, as well as large galaxy surveys. Note that this approach averages over any local effects such as clustering. First, we consider a recent semi-analytic model implemented in the Millennium dark matter simulation (*75*), after re-scaling to the most recent *Planck* cosmological parameters. We extracted mock-observed galaxy catalogs in 24 circular cones with 2 deg. opening angles from the online database (*76*), and inspected the redshift distributions of galaxies with similar magnitudes to the five VHS sources. Although the redshift distributions are wide, each source has between a 5% and 19% probability of having a redshift $z < 0.2$, which is the lower limit on the redshift of VHS7. Lower bounds on the redshifts of each source (95% confidence) are also listed, which range between 0.11 and 0.19. The results are listed in Table S4.

We also inspected the Cosmological Evolution Survey (COSMOS) photometric redshift catalog (*77, 78*). We extracted all galaxies with photometric redshift fits, and measured J and K band magnitudes (we assume equivalence between the K and Ks bands). Based on this catalog, we find consistent results for the probabilities of the VHS sources having redshifts $z < 0.2$, and for the lower bounds on the redshifts. These results are also listed in Table S4.

Finally, we considered the possible redshifts of galaxies fainter than our completeness limit of Ks = 19.2. In both the Millennium-simulation and COSMOS catalogs, we found that >95% of galaxies fainter than Ks = 19.2 are at redshifts $z > 0.2$. This constraint only applies above the galaxy stellar mass limit of the catalogs, which is approximately $10^9$ solar masses.

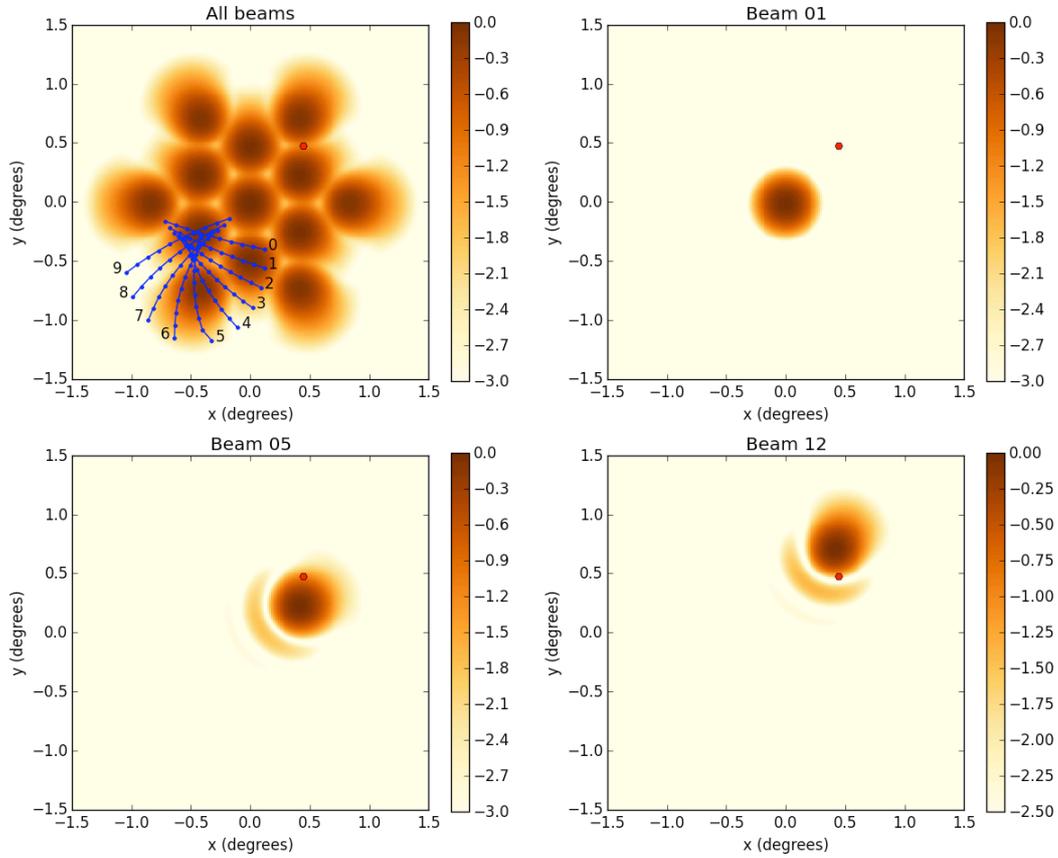

**Fig. S1. Analytic radiation pattern of the MB21 calculated at 1357 MHz.** The top-left panel shows the sum of the radiation patterns of all beams, and the other panels show the radiation patterns of specific beams, as labeled. The color bars represent the base-10 logarithms of the ratios of the intensities at each point to the peak intensities. The blue traces in the top-left panel depict the positions of test pointings; the numbers correspond to the panel labels in Fig. S2. The red circle in each panel represents the best-fit position of FRB 150807 in the Parkes focal plane.

**A.** Beam 2

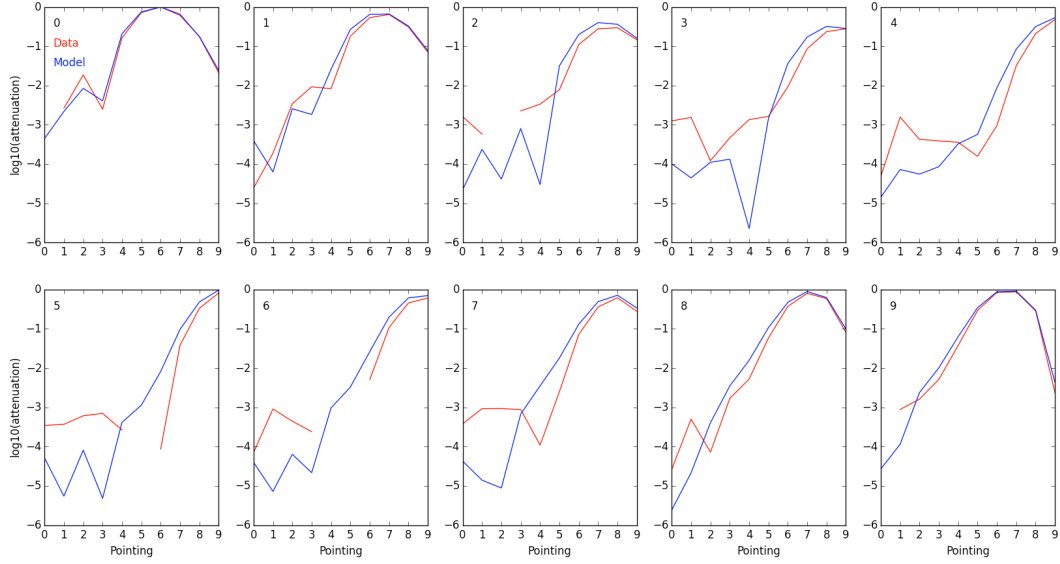

**B.** Beam 9

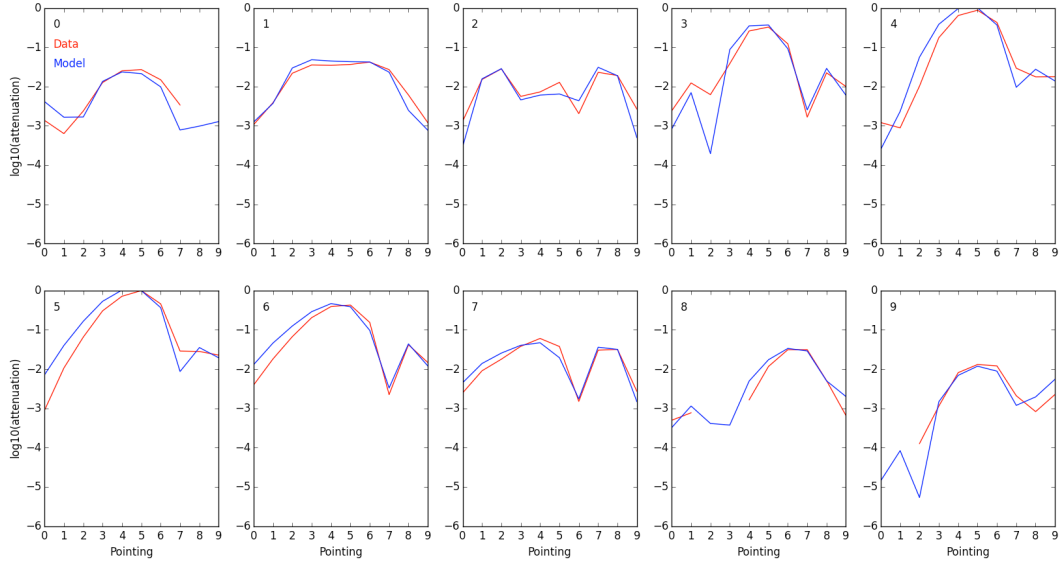

**Fig. S2. Predicted and measured MB21 radiation patterns.** We compare SNRs predicted by the analytic MB21 radiation pattern model (blue traces), and measured using observations of PSR J1644-4559 (red traces). Results for beam 2 are shown in panel A, and results for beam 9 are shown in panel B. Missing measurements indicate non-detections of the pulse, and the noise floor is approximately -3.0 in both cases. Both the predictions and the measurements correspond to the average within a 50 MHz band centered on 1357 MHz. The 10 scans, labeled in the top-left corner of each panel,

correspond to the 10 scans labeled in Fig. S1. Each scan has 10 pointings; pointing 0 in each case is closest to the numerical label in Fig. S1. All measurements are normalized by the SNRs for the pointing of maximum sensitivity in a given beam; this need not be the center of each beam radiation pattern.

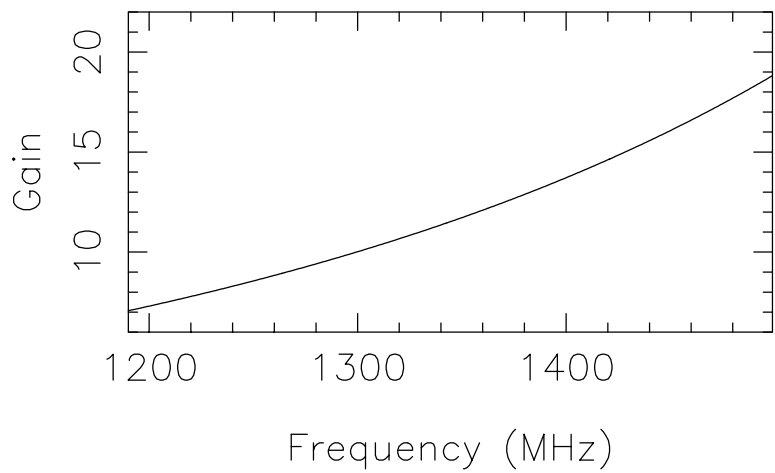

**Fig. S3. Correction factors applied to the measured burst flux densities**. We show the factors applied at different frequencies in beam 5, to account for the off-axis position.

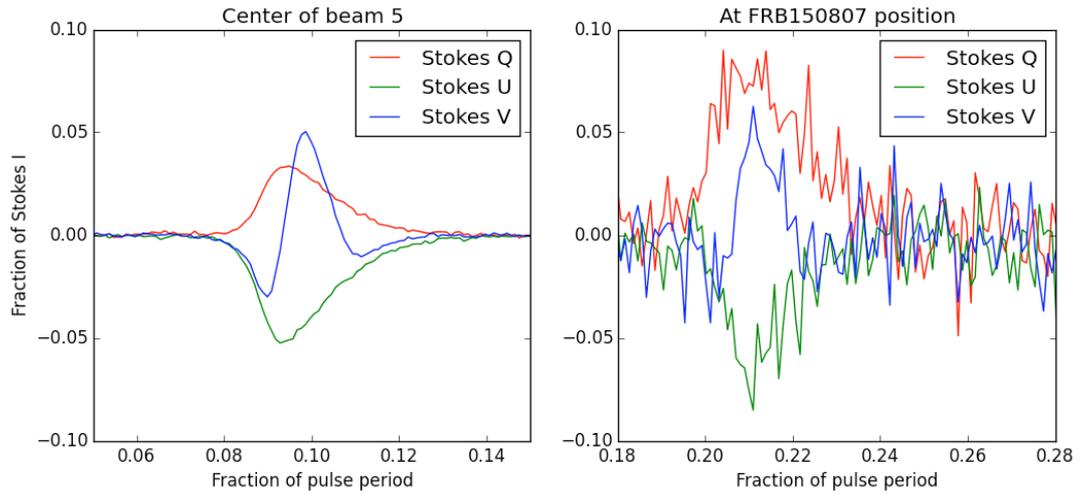

**Fig. S4. Integrated, calibrated pulse profiles in Stokes Q, U, and V of PSR J1644-4559.** The profiles were obtained on-axis in beam 5 of the MB21 (left), and at the offset position of FRB 150807 (right). Each observation was 2 min in length; intrinsic pulse-to-pulse variations will hence add uncertainty to the measurements in addition to thermal noise. The profiles are scaled to the maximum Stokes I measurement.

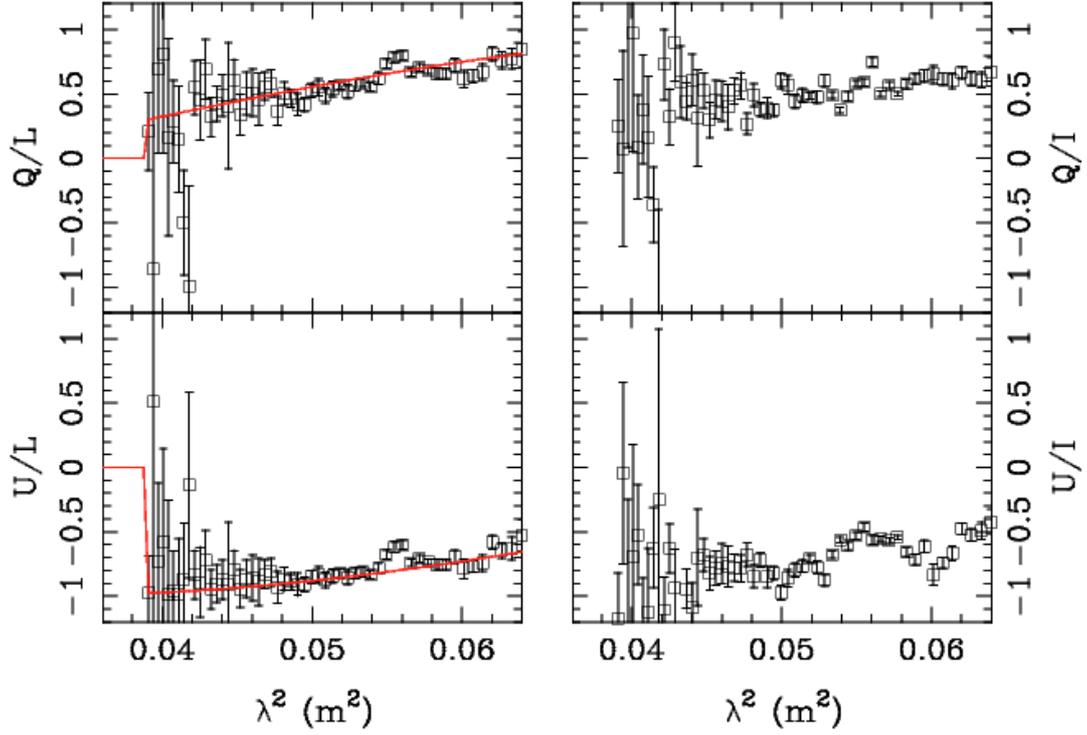

**Fig. S5. Data used to estimate the RM of FRB 150807.** We show pulse-averaged measurements of Stokes Q and Stokes U, normalized by $L = (Q^2+U^2)^{1/2}$ in the left two panels, and by Stokes I in the right two panels. Errors are derived from off-pulse data. The red curves in the left two panels represent the best-fit model for the polarization, including the effects of Faraday rotation. The close match between the left and right panels demonstrates that the pulse is highly linearly polarized, and that the polarized intensity shows similar spectral variations to the total-intensity (Stokes I) spectrum.

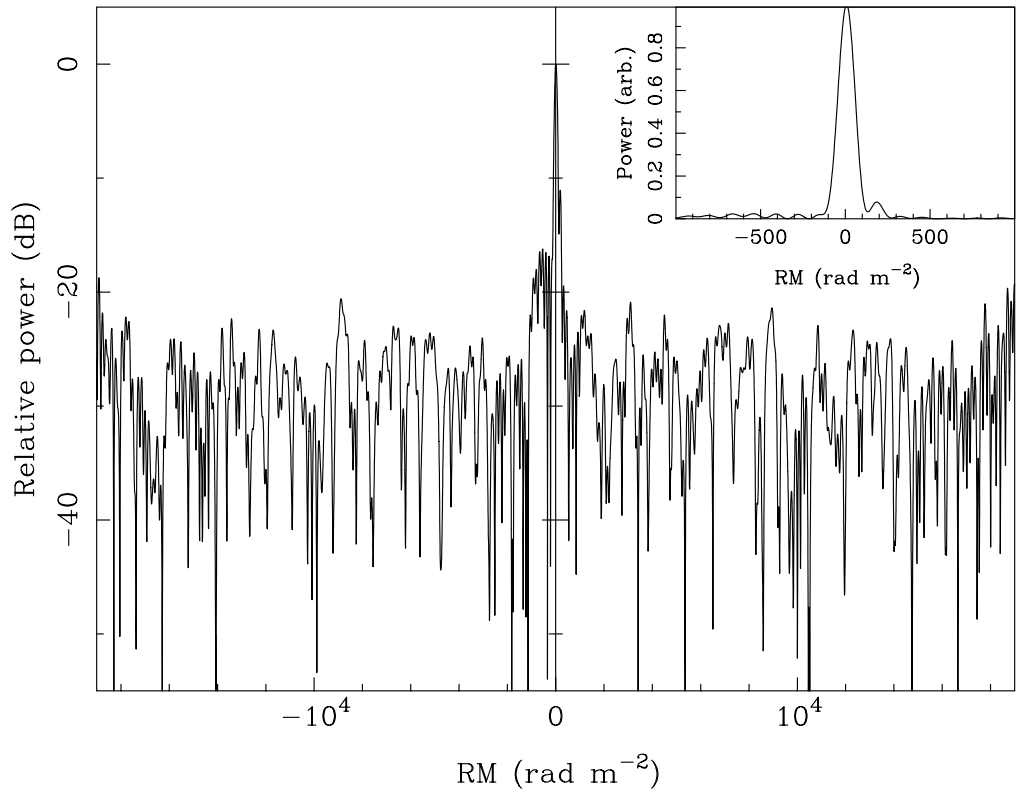

**Fig. S6. Results of our RM synthesis analysis.** The main panel shows the recovered FRB linearly polarized flux density (averaged over the pulse) at different RMs, relative to the peak. The inset shows the central portion of the main panel, with a linear ordinate scale.

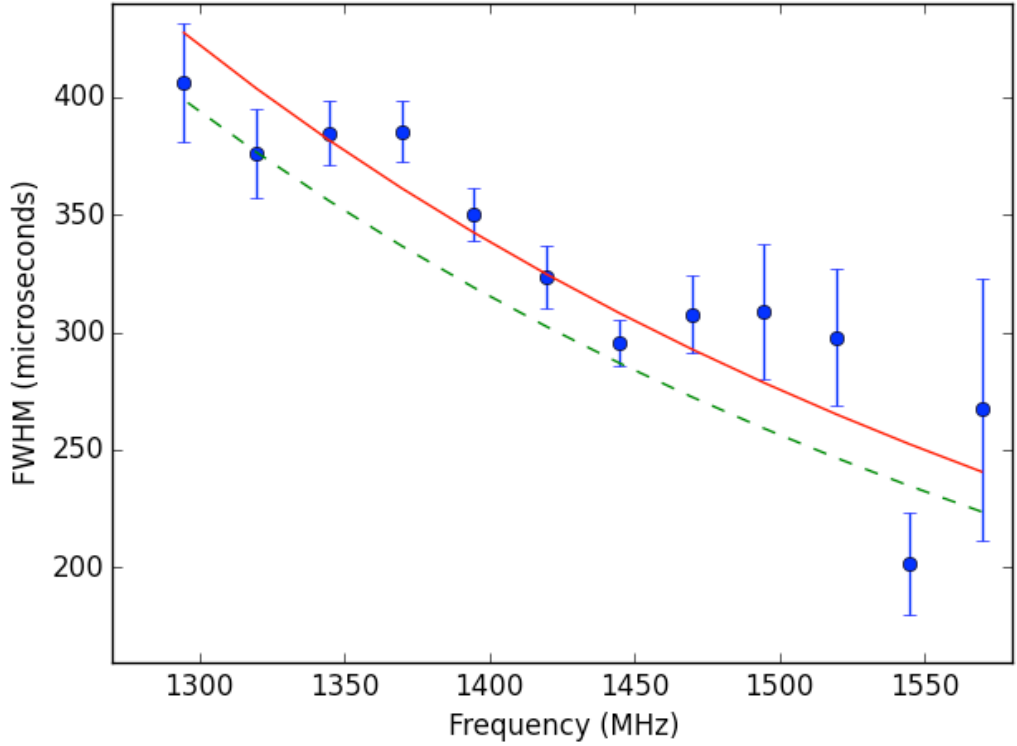

**Fig. S7. Measurements of the FWHM of the FRB 150807 temporal profile (blue points) at different frequencies.** We assume a Gaussian pulse shape. The red line is a fit to these points assuming a power-law form $w_0(f/f_0)^k$, where $w_0$ is the pulse FWHM at a reference frequency $f_0$, and $k$ is the frequency-variability index. We find $k = -3.0\pm0.4$, which is consistent with the pulse profile being determined by dispersion smearing in individual spectrometer channels ($k = -3$ is the expected value in this case). The green dashed line shows the predicted dispersion smearing time-scale across individual channels at each frequency, assuming no overlap in channel frequency-responses.

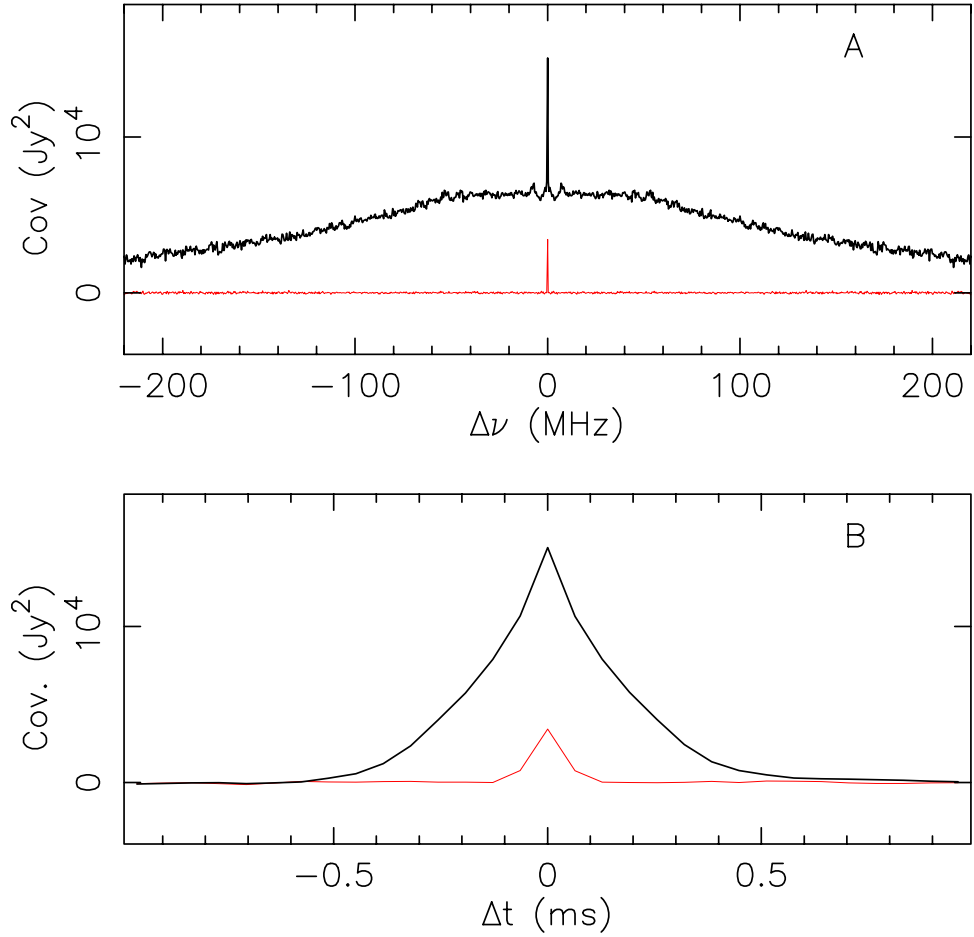

**Fig. S8. Autocovariance functions (ACFs, black) of FRB 150807 in frequency (top, panel A) and time (bottom, panel B).** Prior to calculation of the ACFs, we fully calibrated the data, and subtracted mean levels from each frequency channel. To calculate the frequency-ACF, we averaged the pulse in time across the 7 most significant 64-microsecond bins (indicated by dashed lines in Fig. 1 C), maintaining a 390.625 kHz frequency resolution. To calculate the time-ACF, we averaged the pulse in frequency, maintaining a 64 microsecond time resolution. ACFs of equivalent off-pulse sections of data are shown in red, which are largely consistent with Gaussian noise. The red curves show the covariances derived from an off pulse region of the observation.

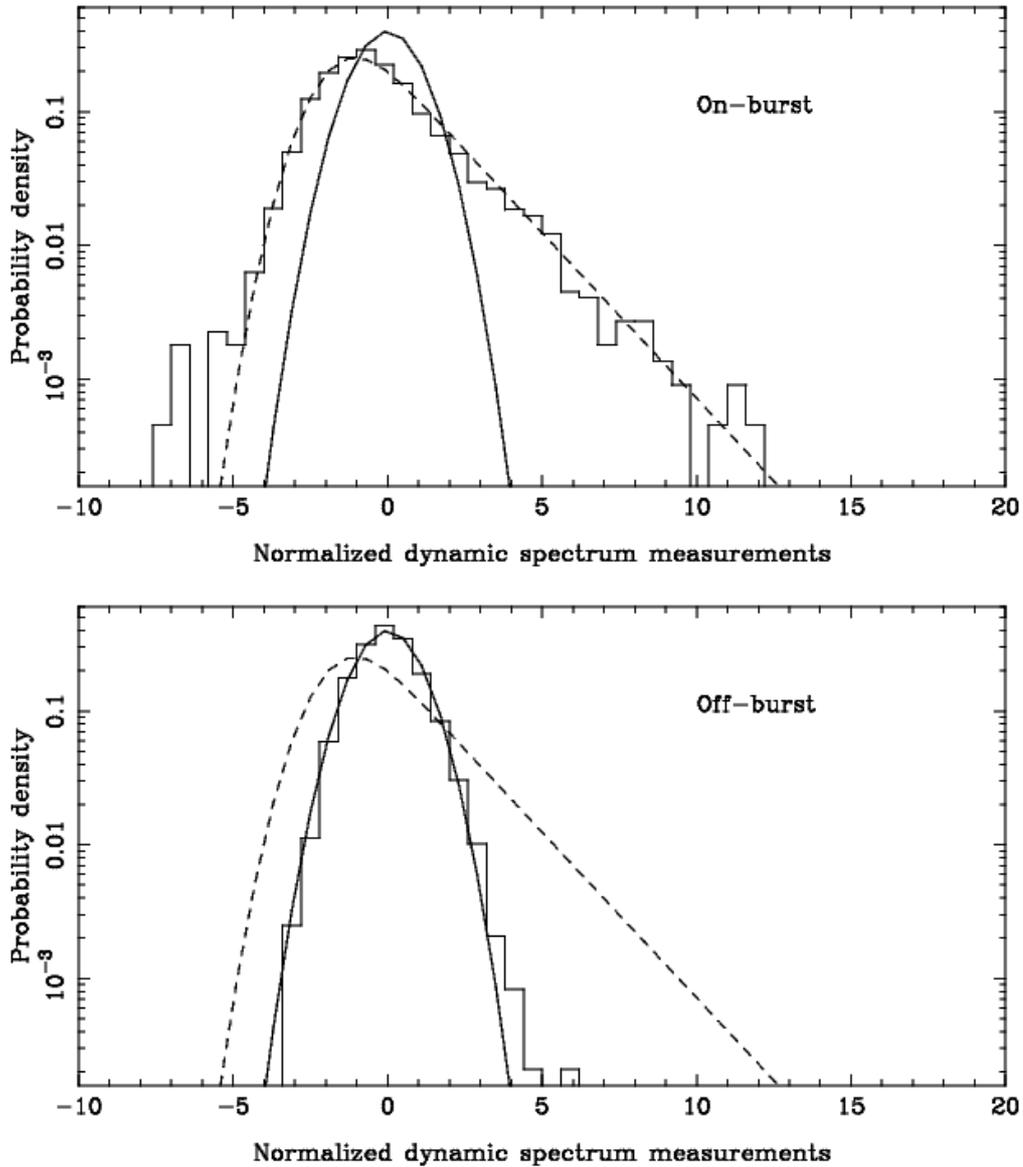

**Fig. S9. Histograms of measurements on (top) and off the burst (bottom) in the dynamic spectrum of FRB 150807.** The histograms were normalized to provide probability density estimates. As described in supplementary section S1, the data were scaled according to the predicted thermal noise level such that, if they were consistent with thermal noise, the distribution would be normal with mean zero and unit variance [$N(0,1)$]. We show such a distribution function as solid curves in each panel. Although the off-burst data are largely consistent with the $N(0,1)$ distribution, with some discrepancies possibly caused by RFI, the on-burst data are not. The dashed curve shows a fitted exponentially modified normal distribution, with rate parameter 0.57 and mean –2.0.

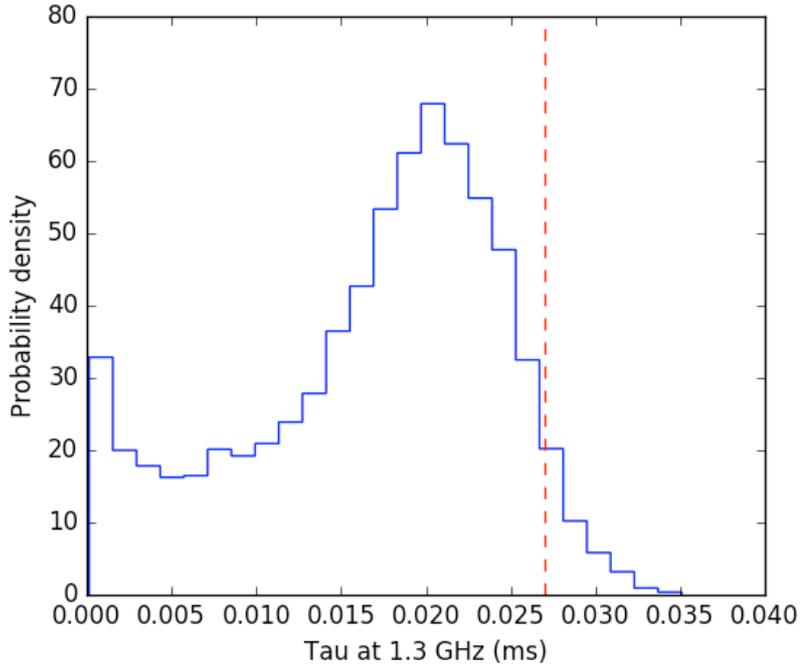

**Fig. S10. Marginalized posterior density estimate for the scattering timescale, $\tau_d$, at 1.3 GHz of FRB 150807 (blue histogram)**. The 95% confidence upper limit of 27 μs is indicated as a red dashed line.

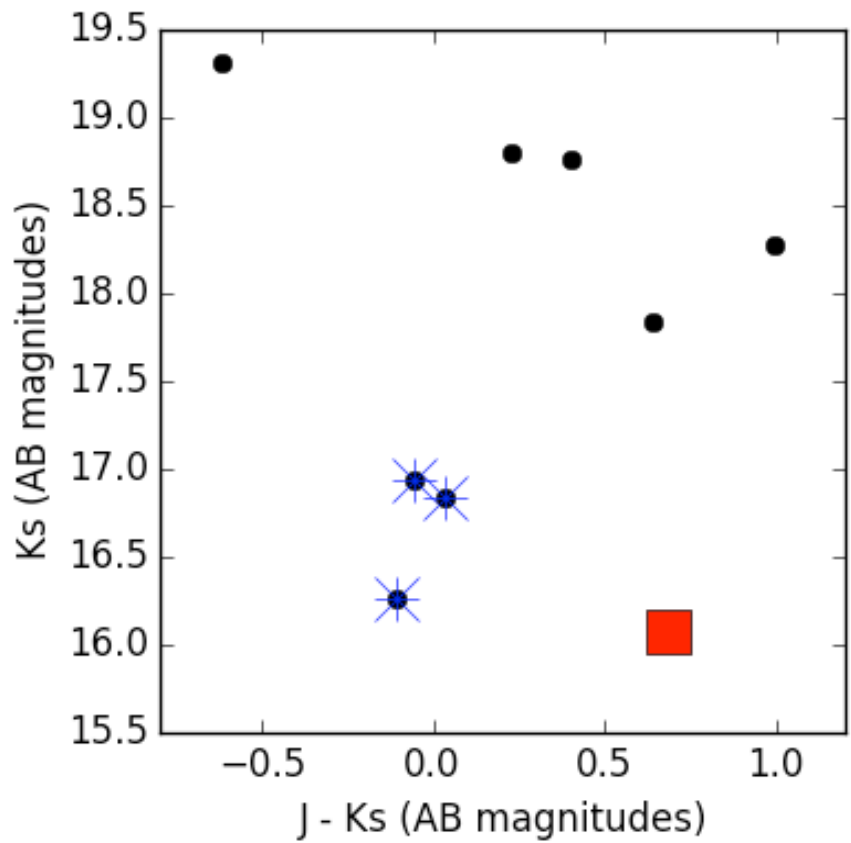

**Fig. S11. Near-IR color-magnitude diagram for sources detected in archival VISTA hemisphere survey (VHS) images of the 95% containment region of FRB 150807.** The points highlighted by blue stars have *SExtractor* CLASS_STAR parameters >0.9, suggesting a >90% probability that they are unresolved and hence consistent with Milky Way stars (*66*) and the point highlighted by a red square has CLASS_STAR < 0.1. The black dots represent the remaining sources.

| Epoch of center of obs. (UT) | On source time (hr) | Array config. | 5.5 GHz beam shape | 5.5 GHz 3σ limit mJy/beam | 7.5 GHz beam shape | 7.5 GHz 3σ limit mJy/beam |
|---|---|---|---|---|---|---|
| 2015 Aug 08, 17:45 | 4.8 | H75 | 119.2×86.0'', 87.9° PA | 0.36 | 80.6×61.4'', -82.7° PA | 0.27 |
| 2015 Aug 11, 18:15 | 5.3 | H75 | 124.5×85.4'', -87.7° PA | 0.30 | 86.4×61.0'', -80.4° PA | 0.30 |
| 2015 Aug 15, 18:20 | 5.7 | H75 | 86.3×74.6'', 84.1° PA | 0.33 | 61.6×53.2'', -75.5° PA | 0.33 |
| 2015 Sep 03, 17:45 | 3.3 | 750B | 35.8×8.5'', 42.7° PA | 0.24 | 27.5×6.3'', 42.3° PA | 0.27 |

**Table S1. Parameters of the ATCA follow-up observations of the FRB 150807 field.** The times on-source were split evenly between 42 pointings spaced by 4.3 arcmin. The beam shape parameters represent the full-width half-maxima and position angles (PAs, East of North) of the two-dimensional Gaussians convolved with the CLEAN results.

| Source | RA (J2000) | DEC (J2000) | J (AB mags) | Ks (AB mags) | CS | w1 (AB mags) | w2 (AB mags) |
|---|---|---|---|---|---|---|---|
| VHS1 | 22:40:15.21 | -53:12:08.8 | 19.16±0.08 | 18.8±0.1 | 0.4 | 17.1±0.1 | 17.0±0.4 |
| VHS2 | 22:40:17.63 | -53:13:25.9 | 19.27±0.09 | 18.27±0.09 | 0.13 | 16.9±0.1 | 16.3±0.2 |
| VHS3 | 22:40:19.54 | -53:14:09.3 | 16.87±0.02 | 16.83±0.03 | 0.97 | 15.81±0.05 | 15.5±0.1 |
| VHS4 | 22:40:25.86 | -53:16:19.2 | 18.47±0.07 | 17.83±0.07 | 0.58 | 16.18±0.06 | 15.8±0.1 |
| VHS5 | 22:40:25.25 | -53:16:33.6 | 19.01±0.08 | 18.8±0.1 | 0.36 | - | - |
| VHS6 | 22:40:24.64 | -53:16:35.6 | 18.69±0.06 | 19.3±0.1 | 0.39 | 17.2±0.1 | 16.6±0.3 |
| VHS7 | 22:40:24.67 | -53:18:05.4 | 16.75±0.02 | 16.06±0.02 | 0.02 | 15.03±0.03 | 14.68±0.06 |
| VHS8 | 22:40:26.51 | -53:18:47.0 | 16.87±0.02 | 16.93±0.03 | 0.98 | 16.16±0.06 | 16.1±0.2 |
| VHS9 | 22:40:30.87 | -53:19:40.5 | 16.14±0.01 | 16.26(2 | 0.98 | 15.37±0.04 | 15.01±0.07 |

**Table S2. Parameters of the nine objects identified in the 95% containment region for FRB 150807 in VHS data, along with WISE magnitudes.** 'CS' corresponds to the *SExtractor* CLASS_STAR parameter, which is the probability that a particular object is unresolved.

| Source | Bj (mags) | UK-R (mags) | UK-I (mags) |
|---|---|---|---|
| VHS3 | 21.3 | 19.1 | 18.1 |
| VHS7 | 21.2 | 18.9 | 18.0 |
| VHS8 | 19.9 | 18.2 | 17.7 |
| VHS9 | 22.7 | - | - |

**Table S3. Magnitudes for the VHS sources with matches in the SuperCOSMOS catalog.**

| Source | Probability of $z<0.2$ (Millennium) | Probability of $z<0.2$ (COSMOS) | $z_{min}$ (Millennium) | $z_{min}$ (COSMOS) |
|---|---|---|---|---|
| VHS1 | 0.09 | 0.09 | 0.14 | 0.15 |
| VHS2 | 0.12 | 0.14 | 0.13 | 0.11 |
| VHS4 | 0.19 | 0.21 | 0.11 | 0.10 |
| VHS5 | 0.09 | 0.09 | 0.14 | 0.15 |
| VHS6 | 0.05 | 0.06 | 0.19 | 0.18 |

**Table S4. Probabilities for each unidentified VHS source to have a redshift $z<0.2$, and 95% confidence lower bounds on the source redshifts.** We show probabilities and redshift limits derived from a semi-analytic model implemented within the Millennium simulation, and from the COSMOS photometric redshift catalog.